\documentstyle[12pt,preprint,aps,prb]{revtex}
\def\nm{\nonumber}
\def\beqa{\begin{eqnarray}}
\def\beq{\begin{equation}}
\def\F{{\cal{F}}}

\def\eeqa{\end{eqnarray}}
\def\eeq{\end{equation}}
\def\lab{\label}    
\def\pa{\partial}
\def\twx{\theta_{\,\widetilde{x}}}
\def\twy{\theta_{\,\widetilde{y}}}
\def\twz{\theta_{\,\widetilde{z}}}
\def\tx{\theta_{x}}
\def\ty{\theta_{y}}
\def\tz{\theta_{z}}
\def\wz{\widetilde{z}}

\def\l{\Lambda}
\def\Del{\Delta}

\begin{document}

\begin{titlepage}    
\thispagestyle{plain}
\pagenumbering{arabic}
\begin{center}
{\Large \bf 
Picard-Fuchs Ordinary Differential Systems in}
\end{center}
\vspace{-7.0mm}
\begin{center}
{\Large \bf $N=2$ Supersymmetric Yang-Mills Theories}
\end{center} 
\normalsize
\begin{center}
{\large Y\H uji Ohta}
\end{center}
\vskip 1.5em
\begin{center}
{\em Research Institute for Mathematical Sciences }
\end{center}
\vspace{-11.0mm}
\begin{center}
{\em Kyoto University}
\end{center}
\vspace{-11.0mm}
\begin{center}
{\em Sakyoku, Kyoto 606, Japan.}
\end{center}
\vspace{-11.0mm}
\begin{abstract}
In general, Picard-Fuchs systems in $N=2$ supersymmetric 
Yang-Mills theories are realized as a set of simultaneous partial 
differential equations. However, if the QCD scale parameter is used as 
unique independent variable instead of moduli, the resulting Picard-Fuchs 
systems are represented by a single ordinary differential equation (ODE) 
whose order coincides with the total number of independent periods. 
This paper discusses some properties of these Picard-Fuchs ODEs. 
In contrast with the usual Picard-Fuchs systems written in terms of moduli 
derivatives, there exists a Wronskian for this 
ordinary differential system and this Wronskian produces 
a new relation among periods, moduli and QCD scale parameter, 
which in the case of SU(2) is reminiscent of scaling relation of 
prepotential. 
On the other hand, in the case of the SU(3) theory, there are two kinds of 
ordinary differential equations, one of which is the equation directly 
constructed from periods and the other is derived from the SU(3) 
Picard-Fuchs equations in moduli derivatives identified with 
Appell's $F_4$ hypergeometric system, 
i.e., Burchnall's fifth order ordinary differential 
equation published in 1942. It is 
shown that four of the five independent solutions to the latter equation 
actually correspond to the four periods in the SU(3) gauge theory and the 
closed form of the remaining one is established by 
the SU(3) Picard-Fuchs ODE. The formula for this fifth solution is a new 
one. 
\end{abstract}
\end{titlepage}


\begin{center}
\section{Introduction}
\end{center}

\renewcommand{\theequation}{1.\arabic{equation}}\setcounter{equation}{0}

It has been recognized that the low energy effective action 
of $N=2$ supersymmetric Yang-Mills theory for any Lie gauge group 
including at most two derivatives and four fermions 
is dominated by a holomorphic function called 
prepotential $\F$. \cite{Sei} Perturbatively, this prepotential is 
a sum of classical part and one-loop contribution, and further 
contributions from higher loop diagrams are excluded by 
the non-renormalization theorem in $N=2$ theory. However, 
$\F$ was expected to be affected by instantons and hence it's 
non-perturbative determination was a longstanding problem. 

In the case of SU(2) gauge theory, i.e., without any quark hypermultiplet, 
Seiberg and Witten \cite{SW1,SW2} showed that the vacuum configuration of 
the $N=2$ action was parameterized by the moduli $u=\langle \mbox{tr}\phi^2 
\rangle$ ($\phi$ is a complex scalar field in the adjoint representation of 
the gauge group) and singularities on this parameter space (moduli space) 
could split into pieces by instanton effect. This instanton corrected moduli 
space is often called quantum moduli space and Seiberg and Witten \cite{SW1,SW2} 
identified the quantum moduli space with the moduli space of a certain 
elliptic curve of genus one. According to their ansatz, 
since the vacuum expectation value $a=\langle \phi \rangle$ and its 
magnetic dual can be also regarded as periods of 
a meromorphic 1-form on the elliptic Riemann surface, 
these periods can be calculated as a linear 
combination of solutions to Picard-Fuchs equations 
(for a historical review and introduction of Picard-Fuchs equation 
in mathematics, see Gray \cite{Gray}). 
Once the periods are calculated, it is immediate to 
obtain the prepotential because of rigid special geometry. 
In this way, Klemm {\em et al.} \cite{KLT} determined the SU(2) 
prepotential. 

Seiberg and Witten's approach to $N=2$ supersymmetric SU(2) 
Yang-Mills theory \cite{SW1,SW2} was extended to other higher rank gauge 
group cases coupled with or without quark 
hypermultiplets, \cite{KLT,KLYT,AF,APS,HO,DS1,DS2,BL,Han,AAM,LPG,AAG} 
and it was found that the quantum moduli spaces of those gauge theories 
could be identified with those of Riemann surfaces with certain 
genus. In these studies, general algorithms to get 
Picard-Fuchs equations were developed, \cite{IMNS,IMNS2,IMNS3,Ali,Ali2} but 
these equations are in general realized as 
a set of simultaneous partial differential equations (PDEs) in terms of 
moduli derivatives. For this reason, it is not easy to solve 
Picard-Fuchs equations, especially, in higher rank gauge goup cases. 

However, if the QCD scale parameter instead of moduli is used as 
unique independent variable, the resulting Picard-Fuchs systems 
will be represented by a single ordinary differential equation (ODE) 
whose order coincides with the total number of independent periods. 
Then the problem solving Picard-Fuchs equations can be encoded into 
the language of ODE and therefore the study of periods are simplified. 
As another feature of this formalism, we remark that 
in contrast with the usual Picard-Fuchs systems written in terms of moduli 
derivatives there exists a Wronskian for these 
ordinary differential systems and the Wronskian produces 
a new relation among periods, moduli and QCD scale parameter. Especially, 
in the case of SU(2) it is quite reminiscent of scaling relation of the 
prepotential. \cite{Mat,STY,EY} This relation is highly non-linear in the 
case of higher rank gauge group, but it reflects the structure of the 
Picard-Fuchs ODE. Sec. II discusses these Picard-Fuchs ordinary differential 
systems. This realization of Picard-Fuchs systems via ODE becomes 
interesting when we consider a 
relation to hypergeometric differential equations in multiple variables. 
For example, in the case of SU(3) gauge theory, we can find another ODE 
which gives equivalent periods. That is Burchnall's fifth order equation 
directly constructed from Appell's $F_4$ hypergeometric differential 
equations.\cite{Bur} In general, the dimension of the solution space of 
a single ODE constructed from simultaneous PDEs 
can exceed that of the original PDEs \cite{Bur} (see also Srivastava and 
Karlsson \cite{SK} and references therein), and Burchnall's equation is 
the case. In addition, the extra solution which is not a solution to 
the original PDEs is known to have a very characteristic form. 
In the case of Burchnall's equation for the SU(3) gauge theory, since four 
of the five independent solutions are found to correspond to the four 
periods of the SU(3) gauge theory (this identification is explicitly 
checked at the semi-classical regime) and these four periods are also 
solutions to the SU(3) Picard-Fuchs ODE, it is possible to extract a 
differential equation only for the fifth solution (although the fifth 
solution is irrelevant to the underlying physics). 
The formula for this fifth solution obtained in this way 
is a new one and takes a very different form compared with other 
fifth order ODEs constructed from a set of PDEs, e.g., 
a product of two Bessel functions or Whittaker functions 
(the ``fifth solution'' to these two cases are quite reminiscent each 
other). In our derivation, it is crucial to notice that 
the fifth solution of the fifth order ODE (that is, Burchnall's equation) 
associated with Appell's $F_4$ is constructed by ``subtracting'' fourth order 
equation satisfied by only periods of the SU(3) Seiberg-Witten curve 
which are part of the solutions to the fifth order ODE. 
In Sec. III, we discuss these aspects of Burchnall's equation as application 
of the SU(3) Picard-Fuchs ODE to a theory of hypergeometric equations. 
Sec. IV is a brief summary. 

{\bf Remark:} {\em When we simply say as 
``Picard-Fuchs ODE'', it always means a single ordinary differential 
equation in terms of QCD scale parameter derivatives.}

\begin{center}
\section{Picard-Fuchs ordinary differential systems}
\end{center}

\renewcommand{\theequation}{2.\arabic{equation}}\setcounter{equation}{0}

\begin{center}
\subsection{The hyperelliptic curve}
\end{center}

Firstly, let us recall that the exact solution to the $SU(n+1)$ 
($n\in \mbox{\boldmath$N$}$) gauge theory as an example. On the affine local 
coordinates $x,y\in \mbox{\boldmath$C$}$, 
the hyperelliptic curve and the 
Seiberg-Witten differential are given by \cite{KLT,KLYT,AF,HO,AAG}   
	\beq
	y^2 =\widetilde{W}_{SU(n+1)}^2 -z
	,\ \lambda_{SW}=\frac{x\pa_x \widetilde{W}_{SU(n+1)}}{y}dx
	,\eeq
where $z=\l_{SU(n+1)}^{2(n+1)}$ and 
	\beq
	\widetilde{W}_{SU(n+1)}=x^{n+1}-\sum_{i=2}^{n+1}s_i x^{n+1-i}
	.\lab{wpoly}
	\eeq
Eq.(\ref{wpoly}) shows the simple singularity 
with moduli $s_i$. This hyperelliptic curve can be 
compactified to a Riemann surface of genus $n$ after addition of 
infinity, hence there must be non-contractible $2n$-cycles on 
this surface and these cycles can be 
chosen as the canonical bases, i.e., $\alpha_i \cap \alpha_j =
\beta_i \cap \beta_j =0, \alpha_i \cap \beta_j 
=-\beta_j \cap \alpha_i =+\delta_{i,j}$. 
Then we introduce the period vector 
	\beq
	\Pi =\left(\begin{array}{c}
	a_{D_i} \\
	a_i \end{array}\right)
	,\eeq
where 
	\beq
	a_i =\oint_{\alpha_i}\lambda_{SW},\ 	
	a_{D_i} =\oint_{\beta_i}\lambda_{SW}
	.\eeq
Below, we often denote moduli as $u\equiv s_2$ and $v\equiv s_3$.

\begin{center}
\subsection{Derivation of Picard-Fuchs ODE}
\end{center}

One of physically interesting behavior of periods is in the weak 
coupling region. Of course, the study of periods can be proceeded by 
using Picard-Fuchs equations, and in general periods are represented 
by series in moduli. However, by some rearrangement they are found to 
have a striking feature in their form. Namely, they can be summarized as 
	\beq
	\mbox{periods} = \mbox{classical part} 
	+ \mbox{instanton corrections}
	.\lab{11}
	\eeq
Eq.(\ref{11}) suggests that it is more convenient 
to construct periods as a series in QCD scale parameter $\l$ rather than 
moduli. In a sense, (\ref{11}) might be implied by 
Ito and Sasakura \cite{IS} in their observation of general form of 
(a certain type of) Picard-Fuchs operators, which is summarized 
schematically as 
	\beq
	L=L_{cl} +\l L_{\l}
	.\eeq
This equality means that the Picard-Fuchs operator $L$ is a sum of 
operator $L_{cl}$ whose kernel is classical periods and some operator 
$L_{\l}$. Once the classical periods are known, 
we can calculate instanton corrected periods, that is, the kernel of $L$, 
by assuming $L_{\l}$ as a perturbation term for small $\l$ 
(at semi-classical regime). However, 
since in the case of other gauge theories with higher rank gauge groups 
Picard-Fuchs system is represented by a set of PDEs, 
such perturbative calculation involves technical problems, 
therefore another method to obtain instanton corrected periods should 
be developed. One of the candidates 
in view of differential equation is to construct 
Picard-Fuchs equations by regarding $\l$ as unique independent variable 
instead of moduli. Then the resulting Picard-Fuchs equations 
will be expressed by an ordinary differential equation. 
Since the QCD scale parameter always appears in any gauge theory 
with or without (massive) hypermultiplets, this formulation is convenient 
when we generalize the method to various gauge theories with any rank 
gauge group.  

Now, let us consider the derivation of this ordinary differential 
equation. In general, $k$-times differentiation of $\lambda_{SW}$ 
over $z$ gives 
	\beq
	\frac{d^k \lambda_{SW}}{d z^k}=\frac{\mbox{polynomial in }x}
	{y^{2k+1}}dx
	,\lab{k}
	\eeq
but the right hand side can be decomposed into a sum of 
Abelian differentials and a total derivative term, if the well-known 
reduction algorithm is used. \cite{IMNS,IMNS2,IMNS3,Ali,Ali2} 
Accordingly, collecting (\ref{k}) for various $k$ can generate a 
differential equation. In addition, since the all independent periods 
should be solutions to this equation, 
the order of the equation must coincide with the 
total number of them and in fact it is determined as $2n$. 
This reduction method is easily confirmed, 
if the Seiberg-Witten curves are hyperelliptic type. 
In this way, we get the Picard-Fuchs ODE in the form  
	\beq
	\left[\frac{d^{2n}}{dz^{2n}}+c_{2n-1}\frac{d^{2n-1}}{dz^{2n-1}}+
	\cdots +c_0 \right]\Pi =0
	,\lab{odes}
	\eeq
where $c_i$ are functions in moduli. 

Also when massive quarks are included, we can obtain a similar ordinary 
differential equation, but in this case mass dependent polynomial 
appears in the denominator of the right hand side of (\ref{k}). 
Reduction of such massive differential was also recognized by 
Marshakov {\em et al.} \cite{MMM} in their construction of massive WDVV 
equations.  

\begin{center}
\subsection{Examples of Picard-Fuchs ODE}
\end{center}

Let us see examples of Picard-Fuchs ODE. The first one is the SU(2) case 
and then the coefficients in (\ref{odes}) are given by 
	\beq
	c_1 =\frac{1}{z},\ c_0 =\frac{1}{16z(u^2 -z)}
	.\eeq
Next, let us consider the SU(3) case. In this case, the coefficients 
are 
	\beqa
	c_0&=&{\frac{-45 \left( 3 z - 4 {u^3} + 27 {v^2} \right)}
  	{2 {z^2} \widetilde{\Delta}_{SU(3)} }},\nm\\
	c_1&=&{\frac{45 \left( 1053 {z^2} - 538 z {u^3} + 40 {u^6} + 
    	3267 z {v^2} - 54 {u^3} {v^2} - 1458 {v^4} \right) }
	{2 {z^2} \widetilde{\Delta}_{SU(3)}}},\nm\\
	c_2&=&\frac{1}{4 {z^2} \widetilde{\Delta}_{SU(3)}}\left[
	445905 {z^3} - 8 \left(4 {u^3}-
	 27 {v^2}\right)^3+{z^2} \left(-217368 {u^3}+734589 {v^2}
	\right)\right.\nm\\
	& &\left.+ 36 z \left( 676 {u^6} - 135 {u^3} {v^2} - 
        29889 {v^4} \right) \right],\nm\\
	c_3&=&\frac{1}{z \widetilde{\Delta}_{SU(3)}}
	\left[76545 {z^3} - 162 {z^2} 
      	\left(244 {u^3}-297{v^2} \right)  - 
     	4 {{\left( 4{u^3}-27{v^2} \right) }^3}\right.\nm\\
	& &\left. +9z\left( 656 {u^6}-1080{u^3}{v^2}- 
        22599{v^4} \right)\right]
	,\lab{211}
	\eeqa
where $\widetilde{\Delta}_{SU(3)}$ is the product
	\beq
	\widetilde{\Delta}_{SU(3)}=(15 z - 4 u^3 + 27 v^2)
	\Delta_{SU(3)}
	\eeq
with the discriminant
	\beq
	\Delta_{SU(3)}=\left[729 z^2 + \left(4 u^3 - 27 v^2\right)^2 - 
        54 z \left(4 u^3 + 27 v^2\right)\right]
	\eeq
of the SU(3) hyperelliptic curve. 

We can easily obtain Picard-Fuchs ODE for SO(5) or Sp(4) gauge group in a 
similar manner. Though the Sp(4) hyperelliptic curve may be seen to be 
different from that of the SO(5) theory, the isomorphism of Picard-Fuchs 
equations between these two theories could be observed by Ito and 
Sasakura. \cite{IS} Also in the case of Picard-Fuchs ODE, 
this isomorphism can be easily established by the same 
transformation.\cite{IS}

\begin{center}
\subsection{Wronskian}
\end{center}

As is well-known, the scaling relation of SU(2) prepotential can be 
generated from Wronskian of Picard-Fuchs equation, but this is valid 
only for this SU(2) theory because Picard-Fuchs equations in other 
gauge theories consist of partial differential equations. In 
such theories, ``Wronskian'' does not generally exist. However, 
our Picard-Fuchs ODE (\ref{odes}) admits a Wronskian given by 
	\beq
	W_{SU(n+1)}=\left|\begin{array}{cccccc}
	a_1 & \cdots & a_n & a_{D_1}&\cdots &a_{D_n}\\
	a_{1}^{\, '}&\cdots &a_{n}^{\, '}&a_{D_1}^{\, '}&\cdots &
	a_{D_n}^{\, '}\\
	\vdots & &\vdots &\vdots & &\vdots \\
	a_{1}^{(2n-1)}&\cdots &a_{n}^{(2n-1)}&a_{D_1}^{(2n-1)}&\cdots &
	a_{D_n}^{(2n-1)}\\
	\end{array}\right|
	,\lab{wronskiod}
	\eeq
where $' =d/dz$. Substituting (\ref{wronskiod}) into (\ref{odes}) shows 
that $W_{SU(n+1)}$ satisfies 
	\beq
	W_{SU(n+1)}^{\, '}+c_{2n-1}W_{SU(n+1)} =0
	,\eeq
which is integrated to give 
	\beq
	W_{SU(n+1)} =\mbox{const. }e^{-\int c_{2n-1}dz}
	,\lab{wwro}
	\eeq
where const. is the integration constant to be determined from the 
comparison of left and right hand sides of (\ref{wwro}) by 
asymptotic behavior of periods, but may depend on moduli 
because in our formulation moduli are regarded as constant. 

Note that (\ref{wwro}) produces a new non-linear relation between periods and other 
parameters. For example, in the case of the SU(2) theory with the 
normalization used by Klemm {\em et al.}, \cite{KLT} (\ref{wwro}) gives
	\beq
	W_{SU(2)}=-\frac{iu}{2\pi z}
	,\eeq
which is a quite reminiscent expression with the homogeneity relation of 
prepotential. \cite{Mat,STY,EY} Similarly, for the SU(3) theory, we have 
	\beq
	W_{SU(3)}=\frac{15z-(4u^3 -27 v^2)}{z^4 \Del_{SU(3)}^2}
	,\lab{su3w}
	\eeq
where the integration constant is normalized to 1 for convenience 
in the next section. Note that the regular singularities of 
(\ref{su3w}) are the same with those of SU(3) Picard-Fuchs ODE. 

\begin{center}
\section{Burchnall's equation}
\end{center}

\renewcommand{\theequation}{3.\arabic{equation}}\setcounter{equation}{0}

\begin{center}
\subsection{Multi-term differential equation}
\end{center}

Picard-Fuchs equations obtained in the previous section 
can be shown to be classified in terms of 
multi-term ordinary differential equation discussed 
by Burchnall in his study of the relationship 
among hypergeometric differential equations in multiple variables and 
certain type of ordinary differential equations.\cite{Bur} 
Here, multi-term ordinary differential equation is defined by:

{\bf Definition :} {\em $k$-term ordinary differential equation 
is the differential equation taking the form 
	\beq
	\left[ f(\tz )+\sum_{i=1}^{l}z^{i}g_i (\tz) \right]\Pi =0
	\lab{15}
	\eeq
for some $l$, where $f$ and $g_i$ are 
polynomial differential operators in the Euler derivative $\tz =zd/dz$. 
$k$ is the total number of $f$ and non-zero $g_i$.}

For example, in terms of the Euler derivative, 
our SU(3) Picard-Fuchs ODE can be rewritten as 
the four-term ODE with
	\beqa
	f&=&2916 {{\left(x-y\right)}^3} 
  	{{\left(-1+{{\theta }_z}\right)}^2}{{{{\theta }_z}}^2},
	\nm\\
	g_1&=&-81\left(x-y\right){{{{\theta}_z}}^2} 
	\left[43x-52y-60\left(x-y\right)  
	{{\theta}_z}+4\left(23x+13y\right)  
        {{{{\theta}_z}}^2} \right],\nm\\
	g_2&=&9\left[\left(386x-251y\right)  
     	{{{{\theta }_z}}^2}+
    	144 \left(2x-7y\right){{{{\theta}_z}}^3}+
    	36 \left(19x+y\right){{{{\theta }_z}}^4}-
    	\left(x-y\right)\left(10+13{{\theta}_z}\right)\right],\nm\\
	g_3&=&-5\left(1+3{{\theta }_z}\right)  
	  \left(2+3{{\theta}_z}\right)\left(-1+6{{\theta }_z} \right)  
	  \left(1+6{{\theta }_z}\right)
	,\eeqa
where $x=4u^3 /27$ and $y=v^2$. 

According to Srivastava and Saran \cite{SS} who extended the work of Burchnall to 
four-term ODE, our SU(3) Picard-Fuchs ODE seems to be representable by a 
hypergeometric function in the homogeneity form $F(pz,qz,rz)$, where 
$p,q$ and $r$ are parameters. In this paper, we could not specify this 
function, but since the kernel of the SU(3) Picard-Fuchs ODE is 
essentially  written by Appell's $F_4$ function, some property of 
$F_4$ may appear in our SU(3) Picard-Fuchs ODE as a four-term equation. 
Furthermore, more detailed study indicates that Picard-Fuchs ODE in any rank 
gauge group can be classified as $k$-term equation, but $k$ seems to 
correspond to $2\times (\mbox{rank of the gauge group})$. 

Finally, note that the Picard-Fuchs ODE in SU(3) gauge theory 
has a factor $(-1+\tz)^2 \tz^2$ in $f$-polynomial. Therefore, the 
indicial indices at semi-classical regime are degenerated to 
$-1$ and $0$. This indicates that there are logarithmic solutions at 
this regime. Of course, similar observation holds also for SO(5) 
and Sp(4) Picard-Fuchs ODEs.

\begin{center}
\subsection{Appell's equations and Burchnall's equation}
\end{center}

As is well-known, a hypergeometric function admits a lot of transformations 
and reducibilities. For example, Gaussian $ _2 F_1 $ system has 24 
solutions and Appell's $F_1$ has 60 solutions. However, these solutions can 
be more systematically constructed, if we consider an equivalent ODE. 
In fact, Srivastava and Saran succeeded to find 120 solutions to the ODE for 
Lauricella's $F_{D}^{(3)}$ function. \cite{SS} 
Also in this sense, study of ODE for hypergeometric partial differential 
system is interesting. The method 
used in these studies followed to Burchnall's work. \cite{Bur} 
In this paper, we do not attempt to obtain all solutions solutions to 
Burchnall's equation (see below) like Kummer's 24 solutions, 
but we can show that the basic five solutions to Burchnall's equation, 
especially, the extra solution which is not a solution to $F_4$ system, can 
be derived by using SU(3) Picard-Fuchs ODE. Of course, it may be interesting 
if also this extra solution can be represented by Appell's $F_4$, but we 
do not know whether it is possible or not. Nevertheless, we can establish 
the fifth solution as a simple formula. The reader should notice that our 
method presented in this paper is to use a fourth order ODE (SU(3) 
Picard-Fuchs ODE) satisfied by periods of Riemann surface in genus two 
(SU(3) Seiberg-Witten curve) and therefore our method is quite different 
from those mentioned above. 

Firstly, let us recall that the SU(3) Picard-Fuchs system \cite{KLT} 
	\beqa
	& &\left[\twx \left(\twx-\frac{1}{3}\right)-\widetilde{x}
	\left(\twx +\twy -
	\frac{1}{6}\right)\left(\twx +\twy -\frac{1}{6}\right)\right]
	\Pi =0,\nm\\
	& &\left[\twy \left(\twy-\frac{1}{2}\right)-\widetilde{y}
	\left(\twx +\twy -
	\frac{1}{6}\right)\left(\twx +\twy -\frac{1}{6}\right)\right]
	\Pi =0
	\lab{ap4}
	,\eeqa
where we have introduced $\widetilde{x}=4u^3 /(27\l_{SU(3)}^6 )$ and 
$\widetilde{y}=v^2 /\l_{SU(3)}^6$, and 
$\theta_x =x\pa/\pa x$ and $\theta_y =y\pa /\pa y$ are Euler partial 
derivatives. In the terminology of previous subsection, (\ref{ap4}) consists 
of two two-term equations and is nothing but the Appell's differential 
system for the type $F_4$ hypergeometric double series
	\beq
	F_4 (\alpha,\beta;\gamma,\gamma^{\,'} ;x,y)=
	\sum_{m.n=0}^{\infty}\frac{(\alpha)_{m+n}(\beta)_{m+n}}
	{(\gamma)_m (\gamma^{\,'})_n}\frac{x^m}{m!}\frac{y^n}{n!}
	,\eeq
where $\alpha=\beta=-1/6, \gamma=2/3$ and $\gamma^{\,'} =1/2$. 
However, by the scaling transformation  
	\beq
	\widetilde{x}=x\wz,\ \widetilde{y}=y\wz,\ 
	x=\frac{4u^3}{27},\ y=v^2 ,\ \wz=\frac{1}{\l_{A_2}^6}
	,\eeq
we see that (\ref{ap4}) turns to 
	\beqa
	& &\left[ \tx (\tx +\gamma -1)-x\wz \,(\tx +\ty +\alpha)
	(\tx +\ty +\beta)\right]F =0,\nm\\
	& &\left[ \ty (\ty +\gamma^{\,'} -1)-y\wz \,(\tx +\ty +\alpha)
	(\tx +\ty +\beta)\right]F =0
	,\lab{F}
	\eeqa
whose one of analytic solutions near $(x,y)=(0,0)$ is given by  
	\beq
	F(x\wz ,y\wz \,)=\sum_{m,n=0}^{\infty}\frac{(\alpha)_{m+n}
	(\beta)_{m+n}}{(\gamma)_m (\gamma ')_n}
	\frac{(x\wz \,)^m}{m!}\frac{(y\wz \,)^n}{n!}
	.\lab{appe}
	\eeq
At first sight, since $F(x\wz ,y\wz \,)$ reduces to $F_4$ 
for $\wz \rightarrow 1$, this scale transformation may be trivial, 
but (\ref{F}) was used as a starting point in the Burchnall's 
work on a set of partial differential equations.\cite{Bur}

In fact, Burchnall noticed on the homogeneity relation of $F$ 
	\beq
	(\tx +\ty -\twz )F=0
	,\eeq
where $\twz$ is the ordinary differential operator $\twz =\wz\, d/d\wz$, and 
finally arrived at the ordinary differential equation of 
fifth order (see also appendix A)  
	\beqa
	& &\left[f_0 -2(x+y)\wz \, f_1 (\twz+\alpha)(\twz+\beta) +
	\frac{1}{2}(x-y)\wz\, f_2 (\twz+\alpha)(\twz+\beta)\right.\nm\\
	& & \left. +(x-y)^2 \wz^{\,2} f_3 (\twz+\alpha)(\twz+\alpha+1)(
	\twz +\beta)(\twz+\beta+1)\right]F=0
	,\lab{fif}
	\eeqa
where 
	\beqa
	f_0&=&\twz\left(\twz+\gamma-1\right)\left(\twz+\gamma^{\,'} -1\right)
	\left(\twz+\gamma+\gamma^{\,'} -2\right)\left(\twz+\frac{\gamma}{2}
	+\frac{\gamma^{\,'}}{2}-2\right),\nm\\
	f_1&=&\left(\twz+\frac{\gamma}{2}+\frac{\gamma^{\,'}}{2}\right)\left(
	\twz +\frac{\gamma}{2}+\frac{\gamma^{\,'}}{2} -\frac{1}{2}\right)
	\left(\twz +\frac{\gamma}{2}+\frac{\gamma^{\,'}}{2}-1\right),\nm\\
	f_2&=&\left(\gamma-\gamma^{\,'}\right)\left(\gamma+\gamma^{\,'}-
	2\right)\left(
	\twz+\frac{\gamma}{2}+\frac{\gamma^{\,'}}{2}-\frac{1}{2}\right),\nm\\
	f_3&=&\left(\twz +\frac{\gamma}{2}+\frac{\gamma^{\,'}}{2}+1\right)
	.\eeqa
In contrast with the SU(3) Picard-Fuchs ODE, since 
Burchnall's equation (\ref{fif}) is classified as a three-term ODE, 
it reflects a consequence of the nature of two variables hypergeometric 
function in the form $F(pz,qz)$, where $p$ and $q$ are 
parameters. \cite{Bur} 

As a direct calculation shows, (\ref{fif}) can not be 
expressed in the form $[\twz +K(x,y,\wz \,)]LF=0$, where 
$K(x,y,\wz \,)$ is some function of $x,y$ and $\wz$, and 
$L$ is some differential operator of fourth order. Accordingly, 
such $L$ does not exist and therefore the relation between our SU(3) 
Picard-Fuchs ODE and Burchnall's equation is non-trivial. 
However, note that the coefficient of highest power in the Euler 
derivative corresponds to the discriminant of the SU(3) curve 
	\beqa
	\Del &\equiv&\left[1-2(x+y)\wz +(x-y)^2 \wz^{\,2}
	\right]\nm\\
	&=&\frac{1}{729\l_{SU(3)}^{12}}\left[ 
	4u^3 -27(v+\l_{SU(3)}^3)^2\right]
	\left[ 4u^3 -27(v-\l_{A_2}^3)^2\right]
	.\eeqa
Therefore, in a sense Burchnall's equation is reminiscent of the SU(3) 
Picard-Fuchs ODE, but these two are not completely equivalent. 
Clarifying the relation between these two equtions is the subject in 
the rest of the paper. 

\begin{center}
\subsection{Four solutions at semi-classical regime}
\end{center}

It would be instructive to get solutions explicitly 
around $\l_{SU(3)}=0$ $(\wz =\infty)$ which is a regular singular point of 
the equation. In the work of Burchnall, solutions of (\ref{fif}) were not 
calculated at any singularities, but since (\ref{fif}) is a linear ordinary 
differential equation, it is easy to solve it by traditional Frobenius's 
method under the assumption $F=\wz^{\,-\nu}\sum_{n=0}^{\infty}
A_n \wz^{\,-n}$ for some $\nu$ and $A_n$. Then the indicial indices are 
determined as  
	\beq
	\nu =\alpha,\ \beta, \ \alpha+1,\ \beta+1,\ 
	\frac{1}{2}(\gamma+\gamma^{\,'} +2)
	\eeq
or equivalently, 
	\beq
	\nu_1 =-\frac{1}{6},\ \nu_2 =\frac{5}{6},\ 
	\nu_3 =\frac{19}{12}
	,\lab{ind}
	\eeq
where $\nu_1$ and $\nu_2$ are actually double roots. The solution 
for $\nu_3$ is the subject in the next subsection.

Eq.(\ref{fif}) produces the recursion relations
	\beqa
	& &(x-y)^2 \rho_1 A_1 -(\nu_i -\alpha)(\nu_i -\beta)\phi_1 A_0 
	=0,\nm\\
	& &(x-y)^2 \rho_n A_n -(\nu_i +n-\alpha-1)(\nu_i +n-\beta-1)
	\sigma_n A_{n-1} +\chi_n A_{n-2}=0
	,\ n>1 
	,\lab{iin}
	\eeqa
where 
	\beqa
	\rho_n &=&(2\nu_{i} +2n-\gamma-\gamma^{\,'}-2)(
	\nu_{i} +n-\alpha)(\nu_{i}+n-\beta)(\nu_{i}+n-\alpha-1)(
	\nu_{i}+n-\beta-1),\nm\\
	\sigma_n &=&(\nu_{i}+n-\gamma-\gamma^{\,'})(2\nu_{i}+2n-
	\gamma-\gamma^{\,'})
	[x(\nu_{i}+n-\gamma^{\,'}-1)+y(\nu_{i}+n-\gamma-1)]\nm\\
	& &+(\nu_{i}+n-1)(2\nu_{i}+2n-\gamma-\gamma^{\,'}-2)[x(\nu_{i}+n-
	\gamma)+y(\nu_{i}+n-\gamma^{\,'})],\nm\\
	\chi_n &=&(\nu_{i}+n-2)(\nu_{i}+n-\gamma-1)(\nu_{i}+n-\gamma^{\,'}-1)
	(\nu_{i}+n-\gamma-\gamma^{\,'})(2\nu_{i}+2n-\gamma-\gamma^{\,'})
	\eeqa
with $A_0 =1 $. Here, repeated indices are assumed {\em not} 
to be summed. If these recursion relations are used, the solutions 
corresponding to respective indicial indices will be 
obtained, but we must be careful, because there are 
indicial indices which differ by unit among them, i.e., 
$\alpha$ and $\alpha +1$, and $\beta$ and $\beta +1$. 
For example, let $\nu =\alpha$. Then the recursion relations produce 
the $n$-th coefficient as a linear combination of 
$A_0$ and $A_1$. Thus the solution is given by a linear 
combination of $\wz^{\,-\alpha}(A_0 +\cdots)$ and $\wz^{\,-\alpha-1}(
A_1 +\cdots)$, but the ``indicial index'' of the 
last series can be seen as $\alpha+1$. Therefore, 
the last series can be also regarded as a solution 
corresponding to this index. In fact, it is easy to see that 
explicit construction of the solution supports this observation. 
Of course, in this case, since we would like to 
get a solution corresponding to the index $\alpha$, 
$A_1$ can be set to zero without loss of generality. 
For this reason, $A_1$ is chosen as zero for 
the indices $\alpha$ and $\beta$, while that for $\alpha +1,\beta+1$ 
and $(\gamma+\gamma^{\,'}-2)/2$ should be determined from 
the first equation in (\ref{iin}). 

In this way, we get the regular series solutions $(i=1,2)$
	\beq
	\varphi_i =\wz^{\,-\nu_i}\sum_{n=1}^{\infty}A_{i,n}\wz^{\,-n}
	,\lab{sersol1}
	\eeq
where the first few coefficients are given by 
	\beqa
	& &A_{i,0}=1,\nm\\
	& &A_{1,1}=0,\nm\\
	& &A_{1,2}={\frac{5}{648 {{\left( x - y \right) }^2}}},\nm\\ 
	& &A_{1,3}={\frac{35 \left(41 x+40 y\right)}
   	{209952 {{\left(x-y\right)}^4}}},\nm\\
	& &A_{2,1}={\frac{5 \left(5 x+4 y\right)}
    	{48 {{\left(x-y\right)}^2}}},\nm\\ 
	& &A_{2,2}={\frac{35 \left(157 {x^2}+460 x y+112 {y^2}
	\right)}{15552 {{\left(x-y\right)}^4}}},\nm\\
	& &A_{2,3}={\frac{385 \left(18671 {x^3}+119352 {x^2} y+
        105504 x {y^2}+12352 {y^3}\right)}{26873856 
	{{\left(x-y\right)}^6}}}
	.\eeqa
On the other hand, the degeneracy of $\nu_1$ and $\nu_2$ produce the 
logarithmic solutions ($j=1,2$)
	\beq
	\widetilde{\varphi}_j =
	\varphi_j \ln\frac{1}{\wz}+\wz^{\,-\nu_j}\sum_{n=1}^{\infty}B_{j,n}
	\wz^{\,-n}
	,\lab{sersol2}
	\eeq
where some of $B_{j,n}$ are 
	\beqa
	B_{1,1}&=&0,\nm\\
	B_{1,2}&=&-{\frac{17}{1296 {{\left(x-y\right)}^2}}},\nm\\
	B_{1,3}&=&-{\frac{\left(11761 x+11216 y\right)}
    	{1259712 {{\left(x-y\right)}^4}}},\nm\\
	B_{2,1}&=&{\frac{49 x+104 y}{144 {{\left(x-y\right)}^2}}}
	,\nm\\
	B_{2,2}&=&{\frac{14273 {x^2}+70940 x y+28268 {y^2}}
 	{46656 {{\left(x-y\right)}^4}}},\nm\\
	B_{2,3}&=&{\frac{41936917 {x^3}+383568144 {x^2} y+
	472854144 x {y^2}+79210112 {y^3}}{161243136 
 	{{\left(x-y\right)}^6}}}
	.\eeqa
It is interesting notice that these series are composed by a series in 
powers of $1/(x-y)=27/(4u^3 -27v^2 )$ which detects the discriminant of the 
semi-classical SU(3) curve. This feature is useful when we consider the 
structure of the quantum moduli space of the SU(3) gauge theory.

For this purpose, let us recall the work of Klemm {\em et al.} \cite{KLT} 
In the course of the analysis of the quantum moduli space of 
the SU(3) gauge theory, Klemm {\em et al.} \cite{KLT} 
found that the quantum moduli space could be better understood 
as the complex projective space $\mbox{\boldmath$CP$}^2$ 
with singularities which correspond to the strong coupling regime. 
Then this space can be covered by the three local (inhomogeneous) coordinates
	\beq
	P_1 :\ \left(\frac{4u^3}{27\l^6}:\frac{v^2}{\l^6}:1\right),\ 
	P_2 :\ \left(\frac{4u^3}{27v^2}:1:\frac{\l^6}{v^2}\right),\ 
	P_3 :\ \left(1:\frac{27v^2}{4u^3}:\frac{27\l^6}{4u^3}\right)
	.\eeq
For this reason, periods should be obtained at each coordinate 
patch, hence the periods derived in this way are locally valid. 
However, our solutions have a (slightly) nice property, because the 
basis of solution space are common both on $P_2$ and $P_3$. That is, 
to get periods on $P_2$, it is enough to further expand $\varphi_i$ 
by $v$, while on $P_3$ by large $u$. 

In fact, we can see that the four periods are expressed by linear 
combinations of (\ref{sersol1}) and (\ref{sersol2}). For example, 
to match periods on the patch $P_3$, let us define $\omega_i$ 
and $\Omega_j$ by linear combinations of $\varphi_i$
	\beqa
	& &\widetilde{\omega}_1 =c_1 \varphi_1 +c_2 \varphi_2 ,\ 
	\widetilde{\omega}_2 =
	c_3 \varphi_1 +c_4 \varphi_2 ,\nm\\
	& &\widetilde{\Omega}_1 =\widetilde{\omega}_1 \ln\frac{27}{4u^3 \wz}
	+\sum_{i=1}^{2}\sum_{n=1}^{\infty}
	c_i B_{i,n}\wz^{\,-\nu_i -n}+c_5 \varphi_1 +c_6 \varphi_2,\nm\\
	& & \widetilde{\Omega}_2 =\widetilde{\omega}_2 \ln\frac{27}
	{4u^3 \wz}+\sum_{i=1}^{2}\sum_{n=1}^{\infty}
	c_{i+2} B_{i,n}\wz^{\,-\nu_i -n}+c_7 \varphi_1 +c_8 \varphi_2,
	,\eeqa
where 
	\beqa
	c_1&=&\ _2 F_1 \left(-\frac{1}{6},\frac{1}{6};\frac{1}{2};
	\frac{27v^2}{4u^3}\right),\ 
	c_2 =-\frac{3}{16u^3}\ _2 F_1 
	\left(\frac{5}{6},\frac{7}{6};\frac{1}{2};\frac{27v^2}{4u^3}
	\right),\nm\\
	c_3&=&\ _2 F_1 \left(\frac{1}{3},\frac{2}{3};\frac{3}{2};
	\frac{27v^2}{4u^3}\right),\ 
	c_4 =\frac{3}{2u^3}\ _2 F_1 
	\left(\frac{4}{3},\frac{5}{3};\frac{3}{2};\frac{27v^2}{4u^3}
	\right),\nm\\
	c_5&=&c_1 +\sum_{n=1}^{\infty}\frac{(-1/6)_n (1/6)_n}{(1/2)_n n!}
	\nm\\
	& &\times\left[\psi\left(n-\frac{1}{6}\right)-\psi
	\left(-\frac{1}{6}\right)+\psi\left(n+\frac{1}{6}\right)
	-\psi\left(\frac{1}{6}\right)\right]\left(\frac{27v^2}{4u^3}
	\right)^n ,\nm\\
	c_6&=&c_2 -\frac{3}{16u^3}\sum_{n=0}^{\infty}
	\frac{(5/6)_n (7/6)_n}{(1/2)_n n!}\nm\\
	& &\times
	\left[2\psi(1)-2\psi(2)+\psi\left(n+\frac{5}{6}\right)
	-\psi\left(-\frac{1}{6}\right)+\psi\left(n+\frac{7}{6}\right)
	-\psi\left(\frac{1}{6}\right)\right]\left(\frac{27v^2}{4u^3}
	\right)^n ,\nm\\
	c_7&=&c_3 +\sum_{n=1}^{\infty}\frac{(1/3)_n (2/3)_n}
	{(3/2)_n n!}\nm\\
	& &\times\left[\psi\left(n+\frac{1}{3}\right)-\psi\left(
	\frac{1}{3}\right)+\psi\left(n+\frac{2}{3}\right)-\psi\left(
	\frac{2}{3}\right)\right]\left(\frac{27v^2}{4u^3}
	\right)^n ,\nm\\	
	c_8&=&c_4+\frac{3}{2u^3}\sum_{n=0}^{\infty}
	\frac{(4/3)_n (5/3)_n}{(3/2)_n n!}\nm\\
	& &\left[2\psi(1)-2\psi(2)+\psi\left(n+\frac{4}{3}\right)
	-\psi\left(\frac{1}{3}\right)+\psi\left(n+\frac{5}{3}\right)
	-\psi\left(\frac{2}{3}\right)\right]\left(\frac{27v^2}{4u^3}
	\right)^n 
	.\eeqa
Here, $\psi (x)=d\ln \Gamma (x)/dx$ is the digamma function and 
$_2 F_1 $ is the hypergeometric function whose series representation 
is given by  
	\beq
	_2 F_1 (a,b;c;x)=\sum_{n=0}^{\infty}
	\frac{(a)_n (b)_n}{(c)_n}\frac{x^n}{n!}
	,\eeq
where $(*)_n =\Gamma (*+n)/\Gamma (*)$ is the Pochhammer symbol. 
In order to make a contact with the normalization used by 
Klemm {\em et al.}, \cite{KLT} we rescale as 
	\beqa
	& &\omega_1 =2\sqrt{u}\l \widetilde{\omega}_1,\ 
	\omega_2 =\frac{v\l}{u}\widetilde{\omega}_2 ,\nm\\ 
	& &\Omega_1 =2\sqrt{u}\l \widetilde{\Omega}_1,\  
	\Omega_2 =\frac{v\l}{u}\widetilde{\Omega}_2  
	.\eeqa
Then the periods $a_j$ and $a_{D_j}$ can be given by 
	\beqa
	& &a_1 =\frac{1}{2}\left( \omega_1 +\omega_2 \right),\ 
	a_2 =\frac{1}{2}\left( \omega_1 -\omega_2 \right),\nm\\ 
	& &a_{D_1}=-\frac{i}{4}\left(\Omega_1 +3\Omega_2 \right)-
	\frac{i}{\pi}\left(\delta_1\omega_1 -\delta_2\omega_2\right)
	,\nm\\ 
	& &a_{D_2}=-\frac{i}{4}\left(\Omega_1 -3\Omega_2 \right)-
	\frac{i}{\pi}\left(\delta_1\omega_1 +\delta_2\omega_2\right)
	,\lab{perio}
	\eeqa
where $\delta_1 =i(5-3\ln 3 -4\ln 2)/4$ and $\delta_2 =3i(1+3\ln 3)/4$ 
are constants determined from asymptotic expansion of 
periods.\cite{KLT} The identification of periods by our solutions can be easily 
established by expanding (\ref{perio}) 
at $u=\infty$. On the other hand, for large $v$, i.e., on the patch $P_2$, 
$\varphi_i$ are expanded at $v =\infty$ 
and then consider similar linear combinations. 

In this way, we can check that the series solutions for $\nu =\nu_1$ 
and $\nu_2$ comprise in fact the four periods. 

\begin{center}
\subsection{The fifth solution}
\end{center}

We have seen that the four solutions with indicial indices $\nu_1$ and 
$\nu_2$ of Burchnall's fifth order equation in fact yield the four periods 
of the SU(3) gauge theory. 
However, there exists an extra solution in (\ref{fif}). Therefore, the 
appearance can be regarded to be characteristic in the ordinary differential 
form of the partial differential system. 

In this subsection, we show that it is possible to derive a fourth 
order equation satisfied by this fifth solution with aid of the SU(3) 
Picard-Fuchs ODE and we derive the closed formula for 
this fifth solution as an application of the SU(3) Picard-Fuchs ODE. 

Firstly, let us rewrite (\ref{fif}) in the form 
	\beq
	\left[ \frac{d^5}{dz^5}+c_{B,4}\frac{d^4}{dz^4}+\cdots +c_{B,0}
	\right] F=0
	,\lab{Burch2}
	\eeq
where $z=1/\wz$ and $c_{B,i}$ are some functions in $x,y$ and $z$. As we 
have already seen in the previous section, since the four of the five 
independent solutions to (\ref{fif}) are regarded as the four 
periods of the SU(3) gauge theory, we can write the Wronskian 
for (\ref{Burch2}) as 
	\beq
	W_B =\left|\begin{array}{ccccc}
	a_1 & a_2 & a_{D_1}& a_{D_2}& h \\
	a_{1}^{\, '} & a_{2}^{\,'} & a_{D_1}^{\,'}& a_{D_2}^{\,'}& 
	h^{\,'} \\
	\vdots & \vdots & \vdots & \vdots & \vdots \\
	a_{1}^{(4)} & a_{2}^{(4)} & a_{D_1}^{(4)}& a_{D_2}^{(4)}& 
	h^{(4)} 
	\end{array}\right|
	,\lab{327}
	\eeq
where $'=d/dz$ and the fifth solution is denoted by $h$. Again from 
basic differential calculation, $W_B$ is generated from $c_{B,4}$ and 
is found to be (c.f (\ref{wwro}))
	\beq
	W_B =\frac{1}{z^{85/12}[x^2 + (y - z)^2 - 2x(y + z)]^3}
	\lab{WronsB}
	\eeq
up to the normalization of integration constant, which is irrelevant 
to the following discussion.

Next, recall that the fourth order derivatives of periods can be reduced 
to a linear combination of lower order derivatives by using the SU(3) 
Picard-Fuchs ODE. Therefore, from (\ref{327}) with the SU(3) Picard-Fuchs 
ODE, we see that 
	\beq
	W_B =W_{SU(3)}(h^{(4)}+c_3 h^{\, '''}+c_2 h^{\, ''}+c_1 h^{\, '}
	+c_0 h )
	,\eeq
where $c_i$ are given by (\ref{211}). 
From (\ref{WronsB}) and (\ref{su3w}), it is immediate to obtain the 
differential equation for $h$
	\beq
	h^{(4)}+c_3 h^{\, '''}+c_2 h^{\, ''}+c_1 h^{\, '}
	+c_0 h=R
	,\lab{heq}
	\eeq
where 
	\beq
	R=\frac{1}{z^{37/12}[5z -9(x-y)][
	x^2 +(y-z)^2 -2 x (y+z)]}
	.\eeq
It is interesting to note that $h$ satisfies the SU(3) Picard-Fuchs ODE 
with a source term. It is now easy to get a general solution to (\ref{heq}) 
(see also appendix B) 
	\beqa
	h&=&\sum_{i=1}^{2}\rho_i a_i +\sum_{i=1}^{2}\epsilon_i a_{D_i}\nm\\
	& &-a_1 \int_0 \frac{Rw_1}{W_{SU(3)}} dx +
	a_2 \int_0 \frac{Rw_2}{W_{SU(3)}} dx 
	-a_{D_1}\int_0 \frac{Rw_3}{W_{SU(3)}} dx +
	a_{D_2}\int_0 \frac{Rw_4}{W_{SU(3)}} dx
	,\lab{hsol}
	\eeqa
where $\rho_i$ and $\epsilon_i$ are integration constants, the 
integration symbol is the integration constant free integral 
and $w_i$ are determinants defined by 
	\beq
	w_1 =\left|\begin{array}{ccc}
	a_2 & a_{D_1} &a_{D_2} \\
	a_{2}^{\,'}&a_{D_1}^{\,'}&a_{D_2}^{\,'}\\
	a_{2}^{\,''}&a_{D_1}^{\,''}&a_{D_2}^{\,''}
	\end{array}\right|,\ 
	w_2 =\left|\begin{array}{ccc}
	a_1 & a_{D_1} &a_{D_2} \\
	a_{1}^{\,'}&a_{D_1}^{\,'}&a_{D_2}^{\,'}\\
	a_{1}^{\,''}&a_{D_1}^{\,''}&a_{D_2}^{\,''}
	\end{array}\right|,\ 
	w_3 =\left|\begin{array}{ccc}
	a_1 & a_2 &a_{D_2} \\
	a_{1}^{\,'}&a_{2}^{\,'}&a_{D_2}^{\,'}\\
	a_{1}^{\,''}&a_{2}^{\,''}&a_{D_2}^{\,''}
	\end{array}\right|,\ 
	w_4 =\left|\begin{array}{ccc}
	a_1 & a_2 &a_{D_1} \\
	a_{1}^{\,'}&a_{2}^{\,'}&a_{D_1}^{\,'}\\
	a_{1}^{\,''}&a_{2}^{\,''}&a_{D_1}^{\,''}
	\end{array}\right|
	.\eeq
In (\ref{hsol}), we have written as $W_{SU(3)}$ for simplicity, but actually 
it should be substituted by (\ref{su3w}). 

To summarize, we have succeeded to find a closed representation of 
the fifth solution by using the SU(3) Picard-Fuchs ODE. It is interesting 
to compare (\ref{hsol}) with the fifth solution of other three-term 
differential equation discussed by Burchnall.\cite{Bur} In the case of a 
product of two Bessel functions, for instance, the fifth solution is expressed 
by an integral of a product of two ``Wronskians'' each of which is written by other 
two independent solutions. \cite{Bur} However, our fifth solution 
does not admit such factorization, so (\ref{hsol}) seems to 
imply the fact that the Appell function $F_4$ with 
parameters $\alpha =\beta =-1/6,\gamma=2/3$ and $\gamma^{\,'}=1/2$ (or 
equivalently, the SU(3) periods) can not be factored into a product of 
two (non-trivial) functions. 

{\bf Remark:} {\em In the semi-classical regime, series representation of 
this fifth solution corresponds to $\nu_3 =19/12$ in (\ref{ind}), and this 
can be also seen from the terms not including $\rho_i $ and $\epsilon_i$ 
in (\ref{hsol}).}

\begin{center}
\section{Summary}
\end{center}

\renewcommand{\theequation}{4.\arabic{equation}}\setcounter{equation}{0}

In this paper, we have discussed the Picard-Fuchs equations appearing 
in $N=2$ supersymmetric Yang-Mills theories in view of ordinary differential 
equation and realized Picard-Fuchs equations as a system of ODEs. 
This construction has given a new 
relation among periods, moduli and QCD parameter by using the Wronskian and 
this is a systematic way to get such non-linear relation among periods in 
higher rank gauge group cases. 

In the case of SU(3), we have also found that 
Burchnall's ordinary differential equation for Appell's $F_4$ is 
a candidate of Picard-Fuchs ODE by identifying the SU(3) QCD mass 
scale parameter with the scaling variable used in Burchnall's observation 
and confirmed that the four of five solutions of Burchnall's equation in 
fact coincide with SU(3) periods. As for the fifth solution, it has been 
shown that it has a simple and closed form by using the SU(3) Picard-Fuchs 
ODE. Of course, even if we consider arbitrary Riemann surface of genus two 
and try to get a similar result for the fifth solution to Burchnall's 
equation, the derivation will be failed because we can not always have an 
equation in fourth order like SU(3) Picard-Fuchs ODE. Note that the SU(3) 
Seiberg-Witten curve is a specific choice and the appearance 
of $\l_{SU(3)}$ plays the central role in the discussion. 

Generalization of Burchnall's construction of ordinary differential equation 
from a set of partial differential equations to Picard-Fuchs equations in 
other gauge theories is straightforward, but we do not know whether 
Burchnall type equations exist or not for Picard-Fuchs equations in those 
gauge theories. Studying these cases will open further aspect of 
Picard-Fuchs equations and hypergeometric nature of the equations 
constructed from the homogeneous hypergeometric equations in multiple 
variables. 

Finally, as another direction, 
since our equation is ODE in contrast with the usual Picard-Fuchs systems, 
it may be possible to consider a relation to 
classical $W$-algebras in view of Picard-Fuchs ODE. \cite{DIZ}

\begin{center}
\section*{Acknowlegment}
\end{center}

I acknowledge Prof. P. W. Karlsson, who kindly gave me a comment 
on my lengthy questions about Burchnall's work. 

\begin{center}
\section*{Appendix A. Derivation of the Burchnall's equation}
\end{center}

\renewcommand{\theequation}{A\arabic{equation}}\setcounter{equation}{0}

In this appendix, we briefly review the derivation of 
the Burchnall's equation. The reader is also recommended to 
refer to the original paper. \cite{Bur}

Firstly, notice that from (\ref{F}) it is easy to obtain 
	\beq
	(\twz +\gamma+\gamma^{\,'} -2)\tx\ty F=(x\wz\, \ty+y\wz\, \tx)XF
	,\lab{eas}
	\eeq 
where $X=(\twz+\alpha)(\twz+\beta)$. We can also obtain 
	\beqa
	UF&\equiv&\left[\twz\, (\twz+\gamma-1)-(x+y)\wz \, X\right]F\nm\\
	&=&\left(2\tx+\gamma-\gamma^{\,'}\right)\ty F,\nm\\
	U'F&\equiv&\left[\twz\, (\twz+\gamma^{\,'}-1)-
	(x+y)\wz \, X\right]F\nm\\
	&=&\left(2\ty+\gamma^{\,'}-\gamma\right)\tx F
	\lab{UU'}
	\eeqa
and 
	\beqa
	& &\tx (\twz+\gamma-1)(\twz+\gamma^{\,'} -1)F=
	\left[(\twz+\gamma^{\,'}-1)\tx\ty+x\wz \, 
	(\twz+\gamma^{\,'})X\right]F,\nm\\
	& &\ty (\twz+\gamma^{\,'}-1)(\twz+\gamma -1)F=
	\left[(\twz+\gamma -1)\tx\ty+y\wz \, (\twz+\gamma)X\right]F
	.\lab{two}
	\eeqa	
Addition of the two equations in (\ref{two}) provides
	\beqa
	VF&\equiv&\left[\twz\, (\twz+\gamma-1)(\twz+\gamma^{\,'}-1)-x\wz\, 
	(\twz+\gamma^{\,'})X-y\wz\, (\twz+\gamma)X\right]F\nm\\
	&=&(2\twz+\gamma+\gamma^{\,'}-2)\tx\ty F
	,\eeqa
thus 
	\beqa
	(\twz+\gamma+\gamma^{\,'} -2)VF&=&\left[4xy\wz^{\,2} XX_{+1}+
	x\wz\, XU+y\wz\, XU' +2(\twz+\gamma+\gamma^{\,'}-2)\tx\ty\right] F
	,\lab{from}
	\eeqa
where $X_{+1}=(\twz+\alpha+1)(\twz+\beta+1)$, and (\ref{eas}) and 
(\ref{UU'}) have been used. Moreover, from (\ref{from}) we have 
	\beqa
	WF&\equiv& \left[\twz\, (\twz+\gamma+\gamma^{\,'}-2)(\twz+\gamma-1)(
	\twz+\gamma^{\,'}-1)\right.\nm\\
	& &-x\wz\left[(\twz+\gamma+\gamma^{\,'}-1)(\twz+\gamma^{\,'})+
	\twz\, (\twz+\gamma-1)\right]X \nm\\
	& &\left.-y\wz\left[(\twz+\gamma+\gamma^{\,'}-1)(\twz+\gamma)+
	\twz\, (\twz+\gamma^{\,'}-1)\right]X +(x-y)^2 \wz^{\,2} 
	XX_{+1}\right]F\nm\\
	&=&2(\twz+\gamma+\gamma^{\,'}-2)\tx\ty F
	.\eeqa
Therefore, the expected equation is given by rearrangement of 
	\beq
	(2\twz+\gamma+\gamma^{\,'}-2)WF=2(\twz+\gamma+\gamma^{\,'}-2)VF
	,\eeq	
i.e., 
	\beq
	\left[Y_0 -x\wz\, Y_1 X-y\wz\, Y_2 X +(x-y)^2 \wz^{\,2} 
	(2\twz+\gamma+\gamma^{\,'}+2)XX_{+1}\right]F=0
	,\lab{ofc}
	\eeq
where
	\beqa
	Y_0&=&\twz\, (\twz+\gamma-1)(\twz+\gamma^{\,'}-1)
	(\twz+\gamma+\gamma^{\,'}-2)
	(2\twz+\gamma+\gamma^{\,'}-4),\nm\\
	Y_1&=&(\twz+\gamma^{\,'})(\twz+\gamma+\gamma^{\,'}-1)
	(2\twz+\gamma+\gamma^{\,'}-2)+
	\twz\, (\twz+\gamma-1)(2\twz+\gamma+\gamma^{\,'}),\nm\\
	Y_2&=&(\twz+\gamma)(\twz+\gamma+\gamma^{\,'}-1)
	(2\twz+\gamma+\gamma^{\,'}-2)+
	\twz\, (\twz+\gamma^{\,'}-1)(2\twz+\gamma+\gamma^{\,'})
	.\eeqa	
Of course (\ref{ofc}) is equivalent to (\ref{fif}).	

\begin{center}
\section*{Appendix B. General solution to fourth order ODE}
\end{center}

\renewcommand{\theequation}{B\arabic{equation}}\setcounter{equation}{0}

This appendix reviews a construction of a general solution to 
the fourth order linear ordinary differential equation 
	\beq
	y^{(4)} +P(x)y^{\, '''}+Q(x)y^{\,''}+R(x)y^{\,'}+S(x)y=T(x)
	,\lab{ki}
	\eeq
where $'=d/dx$ and $P,Q,R,S$ and $T$ are some functions in $x$. 

Firstly, let $T=0$ and let $y_i$ ($i=1,\cdots, 4$) be the fundamental 
solutions to 
	\beq
	y^{(4)} +P(x)y^{\, '''}+Q(x)y^{\,''}+R(x)y^{\,'}+S(x)y=0
	.\eeq
Then we assume that the general solution to 
(\ref{ki}) is represented in the form
	\beq
	y=\sum_{i=1}^{4}C_i (x)y_i 
	\lab{unk}
	\eeq
by using unknown coefficients $C_i$. Differentiating (\ref{unk}), 
we can obtain 
	\beq
	y^{\,'} =\sum_{i=1}^{4}C_{i}^{\,'}y_i +\sum_{i=1}^{4}C_i y_{i}^{\,'}
	,\eeq 
but we further assume that the first term in the right hand side vanishes. 
Namely, we have 
	\beq
	\sum_{i=1}^{4}C_{i}^{\,'} y_{i}=0
	\eeq	
and 
	\beq
	y^{\,'}=\sum_{i=1}^{4}C_{i}y_{i}^{\,'} 
	.\eeq 
Repeating differentiation and imposing the vanishing of terms including 
$C_{i}^{\,'}$, we get 
	\beq
	\sum_{i=1}^{4}C_{i}^{\,'} y_{i}^{\,'}=0,\ 
	\sum_{i=1}^{4}C_{i}^{\,'} y_{i}^{\,''}=0
	\eeq
and 
	\beq
	y^{\,''} =\sum_{i=1}^{4}C_{i}y_{i}^{\,''},\  
	y^{\,'''} =\sum_{i=1}^{4}C_{i}y_{i}^{\,'''}
	.\lab{assu}
	\eeq
As for $y^{(4)}$, we assume 
	\beq
	y^{(4)}=\sum_{i=1}^{4}C_{i}^{\,'}y_{i}^{\,'''}+
	\sum_{i=1}^{4}C_{i} y_{i}^{(4)}
	\lab{b9}
	.\eeq
Then from (\ref{ki}), (\ref{b9}) and (\ref{assu}), we get 
	\beq
	\sum_{i=1}^{4}C_{i}^{\,'}y_{i}^{\,'''}=T
	.\eeq
In this way, we can arrive at the matrix equation determining all $C_i$
	\beq
	YC=\,^{T}(0,0,0,T)
	,\eeq
where 
	\beq
	Y=\left(\begin{array}{ccc}
	y_1 &\cdots & y_4 \\
	y_{1}^{\,'}&\cdots & y_{4}^{\,'}\\
	y_{1}^{\,''}&\cdots & y_{4}^{\,''}\\
	y_{1}^{\,'''}&\cdots & y_{4}^{\,'''}
	\end{array}\right),\ C=\left(\begin{array}{c}
	C_{1}^{\,'}\\
	\vdots\\
	C_{4}^{\,'}\end{array}\right)
	.\eeq
Consequently, $C_i$ are given by 
	\beq
	C_{1}=c_1 -\int_0 \frac{Tw_1}{\det Y} dx,\	
	C_{2}=c_2 +\int_0 \frac{Tw_2}{\det Y} dx,\	
	C_{3}=c_3 -\int_0 \frac{Tw_3}{\det Y} dx,\ 
	C_{4}=c_4 +\int_0 \frac{Tw_4}{\det Y} dx
	,\eeq
where $c_i$ are integration constants, the integration symbol 
is the integration constant free integral and 
	\beq
	w_1 =\left|\begin{array}{ccc}
	y_2 & y_3 &y_4 \\
	y_{2}^{\,'}&y_{3}^{\,'}&y_{4}^{\,'}\\
	y_{2}^{\,''}&y_{3}^{\,''}&y_{4}^{\,''}
	\end{array}\right|,\ 
	w_2 =\left|\begin{array}{ccc}
	y_1 & y_3 &y_4 \\
	y_{1}^{\,'}&y_{3}^{\,'}&y_{4}^{\,'}\\
	y_{1}^{\,''}&y_{3}^{\,''}&y_{4}^{\,''}
	\end{array}\right|,\ 
	w_3 =\left|\begin{array}{ccc}
	y_1 & y_2 &y_4 \\
	y_{1}^{\,'}&y_{2}^{\,'}&y_{4}^{\,'}\\
	y_{1}^{\,''}&y_{2}^{\,''}&y_{4}^{\,''}
	\end{array}\right|,\ 
	w_1 =\left|\begin{array}{ccc}
	y_1 & y_2 &y_3 \\
	y_{1}^{\,'}&y_{2}^{\,'}&y_{3}^{\,'}\\
	y_{1}^{\,''}&y_{2}^{\,''}&y_{3}^{\,''}
	\end{array}\right|
	.\eeq

\begin{center}

\end{center}

\end{document}